\documentclass[aps,prl,reprint,superscriptaddress,longbibliography,floatfix]{revtex4-2}

\usepackage{amsmath,amssymb,bm}
\usepackage{graphicx}
\usepackage{braket}
\usepackage{bbm}
\usepackage[colorlinks=true,linkcolor=blue,citecolor=blue,urlcolor=blue]{hyperref}
\usepackage{booktabs}

\newcommand{\Tr}{\operatorname{Tr}}
\newcommand{\iu}{{i\mkern1mu}}
\newcommand*\diff{\mathop{}\!\mathrm{d}}
\newcommand{\id}{\mathbbm{1}}

\begin{document}

\title{Solvable Random Unitary Dynamics in a Disordered Tomonaga-Luttinger Liquid}

\author{Tian-Gang Zhou}
\email{tgzhou.physics@gmail.com}
\affiliation{Department of Quantum Matter Physics, University of Geneva, 1205 Geneva, Switzerland}

\author{Thierry Giamarchi}
\email{Thierry.Giamarchi@unige.ch}
\affiliation{Department of Quantum Matter Physics, University of Geneva, 1205 Geneva, Switzerland}

\date{\today}

\begin{abstract}
Disordered one-dimensional interacting systems have long been characterized via conventional correlation functions. A complementary quantum-information perspective quantifies the randomness of the unitary ensemble dynamics generated by a quantum system through the frame potential, which serves as a practical diagnostic for quantum algorithmic performance. However, no analytical treatment has yet been achieved for experimentally accessible interacting one-dimensional systems. In this Letter, we derive a closed-form expression for the frame potential of a Tomonaga-Luttinger liquid with quenched Gaussian forward-scattering disorder. Exploiting the exactly quadratic structure of the disorder-averaged Keldysh action, we show that the frame potential decays as a power law at early times and saturates to a late-time plateau controlled by a single coupling parameter. Taking the random field XXZ spin chain as a specific microscopic realization, we show that the strongest randomness is achieved near the Heisenberg ferromagnetic point and can be exponentially enhanced through a multiple-quench protocol. We validate our results across the entire gapless phase, with direct implications for algorithm design in analog quantum simulation platforms.
\end{abstract}

\maketitle

%
\emph{\color{blue}Introduction---}
Disorder in one-dimensional interacting systems has long been a central theme in condensed matter physics. The interplay between interactions and quenched randomness in Tomonaga-Luttinger liquids (TLL) controls transport, spectral, and correlation properties across a wide class of realizations such as organic conductors~\cite{jerome_quasi_2024}, quantum wires~\cite{tarucha_wire_1d,auslaender2005spin}, carbon nanotubes~\cite{bockrath_luttinger_nanotubes} and spin chain materials~\cite{tennant_kcuf_1d}, ultracold atomic gases~\cite{cazalilla_review_bosons} and Josephson junction chains~\cite{fazio_1996,glazman_1997}. Disorder-averaged observables in these settings are traditionally accessed through replica or RG techniques, and the resulting phase diagrams and response functions are by now standard~\cite{GiamarchiSchulz1988,Giamarchi2004}. More broadly, various physical phenomena such as localization in disordered interacting systems have been studied extensively across regimes through these observables~\cite{MBL1,MBL2,MBL3,MBL4}.

A complementary diagnostic for the disordered system has recently emerged from quantum information, namely the proximity of a disordered unitary ensemble to Haar randomness, quantified by the frame potential (FP)~\cite{collinsIntegrationRespectHaar2006a,Scott_2008}. Unlike traditional disorder-averaged correlators that only probe specific observables, the FP characterizes randomness at the level of the unitary ensemble, without reference to specific operators. Such random unitary ensembles generated by local many-body Hamiltonians are known to be essential for certain algorithms~\cite{Vermersch_2019, zhouRealizingUnitarykdesigns2025}, making them a natural target for analysis. So far, however, controlled analytical studies of random unitary ensembles only exist for quantum circuits~\cite{Harrow_2009, Brand_o_2016, Haferkamp2022Quantum, Schuster:2024ajb, PhysRevX.7.021006, Onorati_2017, cui2025randomunitarieshamiltoniandynamics}, all-to-all interacting Hamiltonians~\cite{SachdevYe1993PRL, Kitaev2015Talks, MaldacenaStanford2016PRD, Cotler_2017, Gharibyan:2018jrp}, and Brownian models with stochastic time dependence~\cite{Jian2022LinearGO, Tiutiakina2023FramePO}, but no explicit results exist for these experimentally accessible one-dimensional systems. The obstacle is twofold: theoretically, disorder averaging typically generates nonlocal replica interactions in the field theory, which are difficult to solve analytically. Numerically, approaches are limited by the cost of combining disorder averaging with real-time dynamics at accessible system sizes.

In this Letter, we identify an exactly solvable, experimentally realizable case: a TLL with quenched Gaussian forward-scattering disorder, microscopically realized by the XXZ spin chain with a filtered random field enforcing the forward-scattering condition. In the framework of bosonization, because the disorder linearly couples to the bosonic fields, the disorder-averaged action remains exactly Gaussian, and we obtain a closed-form, nonperturbative expression for the FP. We analyze both the early-time power-law decay and the late-time plateau, which exhibits a one-parameter scaling collapse. We find that the TLL regime near the Heisenberg ferromagnetic point yields the strongest randomness, which is further exponentially enhanced by a multiple-quench protocol~\cite{Vermersch_2019, zhouRealizingUnitarykdesigns2025}. To connect the continuum theory, we benchmark the analytical predictions against exact free-fermion diagonalization and time-evolving block decimation numerics throughout the gapless phase. These results provide deep insight into the random unitary ensemble, with direct implications for algorithm design in analog quantum simulation platforms.

\emph{\color{blue}Model---}
We consider a TLL of length $L$ with Hamiltonian
$H_X = H_{\mathrm{TLL}} + H_{\mathrm{dis},X}$ with $X=U,V$ to be the different random realization index,
where
\begin{equation}
H_{\mathrm{TLL}}=\frac{u}{2\pi}\int \diff x\left[
K(\partial_x\theta)^2+K^{-1}(\partial_x\phi)^2
\right],
\label{eq:HLL}
\end{equation}
$\phi$ and $\theta$ are conjugate bosonic fields,
$u$ is the sound velocity and $K$ the Luttinger parameter~\cite{Giamarchi2004}.
The quenched forward-scattering disorder is
\begin{equation}
H_{\mathrm{dis},X}=\int \diff x\,\xi_X(x)\,\partial_x\phi(x).
\label{eq:Hdis}
\end{equation}
The two disorder realizations $\xi_U$ and $\xi_V$ are independent Gaussian fields with $\overline{\xi_X(x)\xi_Y^*(x')}=\delta_{XY}\,\gamma\,\delta(x-x')$,
where $\gamma$ is the disorder strength.
To investigate the randomness of the unitary evolution, we study the disorder-averaged $k$th FP in terms of Gibbs state
\begin{equation}\label{eq:FP}
    F^{(k)}(T)=\mathbb{E}_{U,V} \left|\Tr\left(\rho W_U(T)W_V^\dagger(T) \right)\right|^{2k},
\end{equation}
via the trace overlap $\mathcal{Z}(T)=\Tr\!\left(\rho W_U(T)W_V^\dagger(T)\right)$, where $\rho=e^{-\beta H_{\mathrm{TLL}}}$ is the thermal density matrix of the clean system, $\beta$ is the inverse temperature. Usually $\rho$ is the maximally mixed state to characterize the unitary $k$-design~\cite{Gross_2007,Roberts_2017}, but here we focus on low temperatures to characterize the randomness. $W_X(T)=e^{-\iu H_XT}$ is the unitary evolution, $T$ being the evolution time.
In a field theory, the absolute normalization of $F^{(k)}$ is nonuniversal. The ratio
$R^{(k)}(T) \equiv F^{(k)}(T)/F^{(k)}(0)$
cancels the UV divergence, and can be computed directly from the path integral. If $W_U$ and $W_V$ are deterministic unitary evolutions, $R^{(k)}$ is identically 1. In the thermodynamic limit, a smaller $R^{(k)}$ indicates a more random ensemble, which is the ultimate goal to incorporate into the algorithm such as randomized measurement~\cite{Vermersch_2019, zhouRealizingUnitarykdesigns2025}. Below, we systematically evaluate this quantity.

\emph{\color{blue}Keldysh formulation---}
Each trace in Eq.~\eqref{eq:FP} admits a Schwinger-Keldysh representation~\cite{Kamenev2011} on a closed contour, and $|\mathcal{Z}(T)|^{2k}$ produces $2k$ Keldysh loops.
For clarity we first describe $k=1$.
The product $\mathcal{Z}(T)\mathcal{Z}^*(T)$ gives two Keldysh loops, each carrying a forward and backward field pair, labeled by field $\Psi=(\phi_U^{+},\phi_V^{-},\phi_V^{+},\phi_U^{-})$ sequentially in the contour.
Performing a loop-wise Keldysh rotation to classical/quantum components and integrating out $\theta$, the clean action becomes~\cite{Giamarchi2004,SM}
\begin{equation}
S_{\rm TLL}
=\frac12\sum_{\omega,q}\Phi^T(q,\omega)\,\bm{G}_0^{-1}(q,\omega)\,\Phi(-q,-\omega),
\label{eq:SLL}
\end{equation}
where $\Phi=(\phi_{1,\rm cl},\phi_{1,\rm q},\phi_{2,\rm cl},\phi_{2,\rm q})$ is the four-component field in loop and Keldysh space. It has the standard upper-triangular structure in each loop, $\bm{G}_0(q,\omega)= \text{Diag}(G_0, G_0)$ where $\text{Diag}$ means the blocked diagonal matrix,
and $G_0 = 
\begin{pmatrix}
    G_0^K & G_0^R \\
G_0^A & 0    \\
\end{pmatrix}$, with the retarded $G_0^R(\omega,q)=\pi K u/\left((\omega+\iu0)^2-(uq)^2\right)$, advanced $G_0^A=(G_0^R)^*$ and Keldysh $G_0^{K}(\omega,q)=\coth\!\left(\frac{\beta\omega}{2}\right)\left(G_0^R-G_0^A\right)$ components of the Green's function.

The disorder effective action from $H_{\text{dis}}$ coupling on the Keldysh contour $C_X$ can be expressed as
\begin{equation*}
    \iu S_{\text{dis}}
	= - \iu \int dx\int_0^{T} dt\, \sum_{X=U,V} \xi_X(x)
	\big[\partial_x\phi_X^+(x,t) - \partial_x\phi_X^-(x,t)\big],
\end{equation*}
with the unrotated field from $\Psi$. Performing the disorder average and the Keldysh rotation, the effective disorder action becomes
\begin{equation}
	\begin{split}
		\iu S^{\text{eff}}_{\text{dis}} &= -\gamma \sum_{q,\omega,\omega'}
		q^2 g_T(\omega) \Phi^T(q,\omega)
		C_0 \Phi(-q,-\omega')g_T(-\omega'), 
	\end{split}
\end{equation}
where $\bm{C}_0(q)\equiv -\iu \gamma q^2 C_0$ is a $4\times4$ matrix in loop and Keldysh space encoding how $U$ and $V$ disorder couple across the two loops. The bare matrix $C_0$ is given in the Supplemental Material~\cite{SM}.
In frequency space the kernel has a rank-one structure controlled by the window function $g_T(\omega)=\omega^{-1} \left(1-e^{\iu\omega T} \right)$.

The full effective action $S=S_{\text{TLL}}+S^{\text{eff}}_{\text{dis}}$ takes the form $S = \frac{1}{2}\sum_{q,\omega,\omega'} \Phi^T(q,\omega)\,\bm{\Omega}(q;\omega,\omega';T)\,\Phi(-q,\omega')$, with:
\begin{equation}
\bm{\Omega}(q;\omega,\omega';T)
=\bm{\Omega}_0^{-1}(q;\omega,\omega')
+g_T(\omega)\,\bm{C}_0(q)\,g_T(\omega'),
\label{eq:Kkernel}
\end{equation}
where $\bm{\Omega}_0^{-1}(q;\omega,\omega')=(2\pi)\delta(\omega+\omega')\bm{G}_0^{-1}$ is the TLL kernel.

Because the action is quadratic, each momentum mode contributes independently and the normalized FP ratio reduces to a product of ratios of functional determinants $R^{(1)}(T)
=
\prod_q
\left(
\frac{\det \bm{\Omega}(q;T)}{\det \bm{\Omega}(q;0)}
\right)^{-1/2}$.
For a system with open boundary conditions (OBC) the momentum is $q = \pi n/(L+1)$ with $n=1,\dots, L$. The separable frequency structure of \eqref{eq:Kkernel} leads, using 
the matrix determinant lemma~\cite{Harville1997}, to
\begin{equation}
    \det\left(\frac{\bm{\Omega}(q;T)}{\bm{\Omega}_0^{-1}(q;T)}\right)
=
\det\!\left(\id_4+\bm{C}_0(q)\,\bm{\Pi}(q;T)\right),
\label{eq:detlemma}
\end{equation}
where the frequency space is internally evaluated with $\bm{\Pi}(q;T)=\int\frac{\diff\omega}{2\pi}\,g_T(-\omega)\,\bm{G}_0(q,\omega)\,g_T(\omega)$.
Since $\det(\bm{\Omega}_0^{-1}(q;T))$ cancels with the denominator in the FP ratio at $T=0$, all nontrivial time dependence arises from $\det(\id_4+\bm{C}_0(q)\bm{\Pi}(q;T))= \ln(1 + A_q(T))$, where
\begin{equation}
   A_q(T) =
\frac{g}{|q|}
\sin^2\!\left(\frac{u|q|T}{2}\right)
\coth\!\left(\frac{\beta u|q|}{2}\right).
\label{eq:AqT} 
\end{equation}
Here $g \equiv 8\pi K \gamma/u^2$ is a dimensionless coupling.


The previous derivation is for $k=1$. The generalization to arbitrary $k$ requires $2k$ replica contours. As shown in the Supplemental Material~\cite{SM}, the result becomes
\begin{equation}
\ln R^{(k)}(T)
=
-\frac{1}{2}\sum_{q}\ln\!\Big[1+kA_q(T) \Big]e^{-\alpha |q|},
\label{eq:main_result}
\end{equation}
where $\alpha$ is the ultraviolet cutoff in momentum space inherited from the bosonization framework~\cite{Giamarchi2004}, and the momentum is $q=\pi n/(L+1)$ with $n=1,\dots,L$. In the thermodynamic limit $L\to \infty$, one replaces $\sum_{q}\to\frac{L}{\pi}\int_0^\infty \diff q$. Eqs.~\eqref{eq:AqT} and~\eqref{eq:main_result} constitute our central result, nonperturbative in $\gamma$ and resumming all orders in the disorder strength. Below we analyze two limiting regimes and identify the optimal parameters for maximizing randomness.


\paragraph{Short-time behavior.}
When $kA_q(T)\ll 1$ for the contributing modes, one can expand the logarithm in Eq.~\eqref{eq:main_result}.
Restricting further to low temperature ($\beta u/\alpha\gg 1$) and weak disorder ($\gamma\ll 1$), so that $\coth(\beta u q/2)\approx 1$. The result reads $ R^{(k)}(T) \simeq
 \left(\frac{\alpha^2+u^2T^2}{\alpha^2}\right)^{-kLg/(8\pi)}$.
It is directly analogous to the correlator in the TLL, where the power-law decay involves dependence on both $K$ and $u$. This is expected, since the random potential induces increasing randomness at later times.



\paragraph{Late-time behavior.}
However, the FP does not decay monotonically but instead saturates at late times, characterizing the maximal randomness of the unitary ensemble. For $uT\gg \alpha$, the factor $\sin^2(uqT/2)$ oscillates rapidly and self-averages under the continuum integral. In this late-time regime, Eq.~\eqref{eq:main_result} can be approximated as
\begin{equation*}
 \ln R^{(k)}(\infty)
\approx
-\frac{L}{4\pi^2}\int_0^\infty \diff q\int_0^{2\pi}\diff{\theta}\,e^{-\alpha q}\,
\ln\!\left(1+\frac{gk}{q}\sin^2(\theta)\right).
\label{eq:Idef}
\end{equation*}
Here $\theta=u|q|T/2$ is treated as a uniformly distributed phase in the long-time limit, replacing $\sin^2(uqT/2)$ by its phase average. Performing the $\theta$ and $q$ integrals analytically and taking the thermodynamic limit $L\to \infty$ first followed by $T\to \infty$, we obtain
\begin{equation}
\ln R^{(k)}(\infty)
=
-\frac{L}{2\pi\alpha}\left(
\ln\!\left(\frac{\alpha gk}{4}\right)
+e^{\alpha gk/2}K_0\!\left(\frac{\alpha gk}{2}\right)
+\gamma_E
\right),
\label{eq:Rinf_closed}
\end{equation}
where $K_0$ is the modified Bessel function of the second kind and $\gamma_E$ is the Euler-Mascheroni constant. 

This formula provides the theoretical framework for identifying the optimal parameter regime to enhance randomness. The plateau value $R^{(k)}(\infty)$ decreases monotonically with $g \equiv 8\pi K\gamma/u^2$. Physically, stronger disorder $\gamma$, a larger Luttinger parameter $K$, and a smaller sound velocity $u$ all enhance the intrinsic randomness of the unitary dynamics. Physically, a smaller sound velocity $u$ means the system evolves more slowly, giving disorder more time to act. Furthermore, the combination $K/u$ is proportional to the compressibility $\kappa$ of the TLL, so $g \propto \kappa \gamma/u$ has a transparent physical meaning: randomness is enhanced when the system is both highly compressible and subject to strong disorder. In particular, a larger $K$ reflects dominant phase fluctuations and greater susceptibility to perturbations. Therefore, to achieve stronger randomness, the system $H_0$ should approach the ferromagnetic limit $K \to \infty$ and $u \to 0$.

\emph{\color{blue}Multiple-quench protocol---}
Although we have analyzed the minimal achievable FP for various parameter values, a generic local spin chain Hamiltonian approaches the Haar ensemble parametrically more slowly than all-to-all models such as the random spin model~\cite{Suter06,Suter07,liEmergentUniversalQuench2024,liErrorResilientReversalQuantum2026,Lukin17NV,Lukin18NV,Hazzard:2014bx,Yan:2013fn,Bloch:2012ee,smaleObservationTransitionDynamical2019,Signoles:2019us} or the Sachdev-Ye-Kitaev model~\cite{SachdevYe1993PRL,Kitaev2015Talks,MaldacenaStanford2016PRD,MaldacenaStanfordYang2016PTEP,BagretsAltlandKamenev2016NPB,GuQiStanford2017JHEP,SongJianBalents2017PRL,Cotler_2017, Gharibyan:2018jrp}. One strategy to enhance the randomness of the unitary ensemble is the multiple-quench protocol, in which the total evolution is a concatenation of evolutions under independently drawn Hamiltonians~\cite{zhouRealizingUnitarykdesigns2025, Vermersch_2018}. Except for the special case of GUE Hamiltonians~\cite{Cotler_2017,zhouRealizingUnitarykdesigns2025}, no analytic results exist for such protocols in general Hamiltonian systems. The quadratic structure of the forward-scattering model, however, allows us to generalize the single-quench result to a sequence of $m$ independent quenches and obtain an analytical formula.
For each realization $X=U,V$, we define
\begin{equation}
W_X(\{T_j\})
=
e^{-iH_{X,m}T_m}\cdots e^{-iH_{X,2}T_2}e^{-iH_{X,1}T_1},
\label{eq:Wm_def}
\end{equation}
where $T_j$ is the evolution time of the $j$th quench and each Hamiltonian $H_{X,j}=H_{\rm TLL}+H_{{\rm dis},X,j}$
contains an independent forward-scattering disorder realization labeled by the quench index $j$, with $\overline{\xi_{X,j}(x)\xi_{Y,\ell}(x')}=\gamma\,\delta_{XY}\delta_{j\ell}\delta(x-x')$.
The FP and its normalized ratio $R_m^{(k)}(\{T_j\})\equiv F_{m}^{(k)}(\{T_j\})/F_{m}^{(k)}(\{0\})$ are defined as in Eq.~\eqref{eq:FP} with $W_X(T)\to W_X(\{T_j\})$.

The derivation follows the same Keldysh route as in the single-quench case, except that the time window is now a sum of segment windows. Deferring details to the Supplemental Material~\cite{SM}, the final result reads
\begin{equation}
\ln R_m^{(k)}(\{T_j\})
=
-\frac{1}{2}\sum_q\sum_{j=1}^m
\ln\!\bigl[1+kA_q(T_j)\bigr]e^{-\alpha |q|}.
\label{eq:main_result_mq}
\end{equation}
Thus, for independent quenches, the contributions simply accumulate in the normalized FP, with the same mode-resolved structure $A_q(T)$ as in the single-quench problem. The late-time plateau accordingly takes the analytical form $\ln R^{(k)}_{m}(\{\infty\}) = m \ln R^{(k)}(\infty)$
where we again take the thermodynamic limit $L\to \infty$ first and then $T_j \to \infty$. This formula shows that the FP is exponentially suppressed with quench order $m$, confirming the effectiveness of the multiple-quench protocol~\cite{Mele2024introductiontohaar,zhouRealizingUnitarykdesigns2025,zhou2026hamiltonianssufficientunitarykdesign}.

\emph{\color{blue} Microscopic model---}
 To test these predictions against a concrete lattice system, we study the spin-$\tfrac{1}{2}$ XXZ chain with open boundary
  conditions 
  and random longitudinal fields,
  \begin{equation*}
    H = J\sum_{i=1}^{L-1}\!\Bigl(S_i^x S_{i+1}^x + S_i^y S_{i+1}^y
        + \Delta\, S_i^z S_{i+1}^z\Bigr) + \sum_{i=1}^{L} \tilde{h}_i S_i^z,
    \label{eq:H_XXZ}
  \end{equation*}
  We set $J=1$. In the gapless regime $|\Delta| < 1$, the clean chain flows to a
  TLL fixed point~\cite{Giamarchi2004,PhysRevLett.45.1358}.  The velocity $u$ and
  Luttinger parameter $K$ are determined exactly by the Bethe ansatz~\cite{Luther1975,PhysRevLett.45.1358,Giamarchi2004}, with
  $\eta = \arccos\Delta$, $K = \frac{\pi}{2(\pi - \eta)}$ and $u = \frac{J\pi\sin\eta}{2\eta}$.
  
  To enforce the forward-scattering approximation~\cite{GiamarchiSchulz1988, Giamarchi2004}, the
  random fields couple to the smooth density mode of the TLL through the
  bosonization identity $S_i^z \approx -(1/\pi)\partial_x\phi(x_i)$,
  neglecting
  rapidly oscillating $2k_F$ components that are irrelevant at long
  wavelengths.
  The relevant low-energy disorder amplitude at each site is the filtered field
  $\tilde{h}_i = (h_{i-1} + h_i)/2$, where $h_i \sim \mathcal{N}(0,\sigma_h^2)$ are i.i.d.\ Gaussian variables, with $\mathcal{O}(1/L)$ boundary corrections.
  The disorder power spectrum, defined as $S(q) \equiv \sum_r e^{-\iu qr}\overline{\tilde{h}_i \tilde{h}_{i+r}} $, is evaluated to be $\cos^2(q/2)\,\sigma_h^2$.
  At $q=\pi$ the spectrum vanishes, suppressing backward scattering, while at long wavelengths $S(0) = \sigma_h^2$. Since the disorder couples via $\sum_i \tilde{h}_i S_i^z$ and the bosonization identification gives $S^z(x) \simeq -(1/\pi)\partial_x\phi(x)$, the continuum forward-scattering correlator is
$\overline{\xi(x)\xi(x')} = \gamma\,\delta(x-x')$ with $\gamma = \sigma_h^2/\pi^2$.
Higher-order filters further suppress backward scattering, as described in the Supplemental Material~\cite{SM}. 


\begin{figure}[t]
\centering
\includegraphics[width=0.95\linewidth]{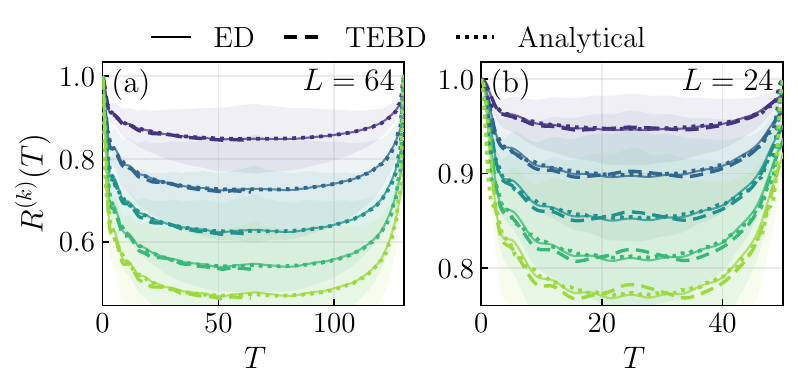}
\caption{
Free-fermion benchmark at $\Delta=0$ with $\sigma_h=0.05$. In each panel, curves for $k=1,\dots,5$ are shown from top to bottom. (a)~$L=64$.
Solid lines are free-fermion numerics from exact diagonalization.
Dashed lines denote TEBD benchmark data, available up to $T=64$.
Dotted lines are the field-theory prediction using the discrete-momentum formula in Eq.~\eqref{eq:main_result}. (b)~$L=24$ with all other settings unchanged. In both cases, a single UV cutoff $\alpha$ is obtained from a $k=1$ and then kept fixed for all $k$.
}
\label{fig:freefermion}
\end{figure}
For all numerics below, we only consider the ground state with $\beta\to \infty$. At $\Delta=0$ the model maps to free fermions, allowing exact diagonalization (ED) to capture long times. All time-evolving block decimation (TEBD)~\cite{Vidal2003, Schollwock2011} simulations use a bond dimension $\chi=256$, a Trotter decomposition with time step $\delta T=0.05$, and $N_{\rm dis}=500$ disorder realizations. As shown in Fig.~\ref{fig:freefermion}, the ED and TEBD results agree for both system sizes. The FP ratio first decays and then revives to unity, a characteristic finite-size effect. The revival occurs when $u|q|T_r/2 = \pi$ in Eq.~\eqref{eq:AqT}, giving a minimal revival time $T_r \approx 2L/u$ with $u=J$ at $\Delta=0$, i.e.\ $T_r\approx 128$ for $L=64$ and $T_r\approx 48$ for $L=24$, consistent with the numerics.

\begin{figure}[t]
\centering
\includegraphics[width=0.95\linewidth]{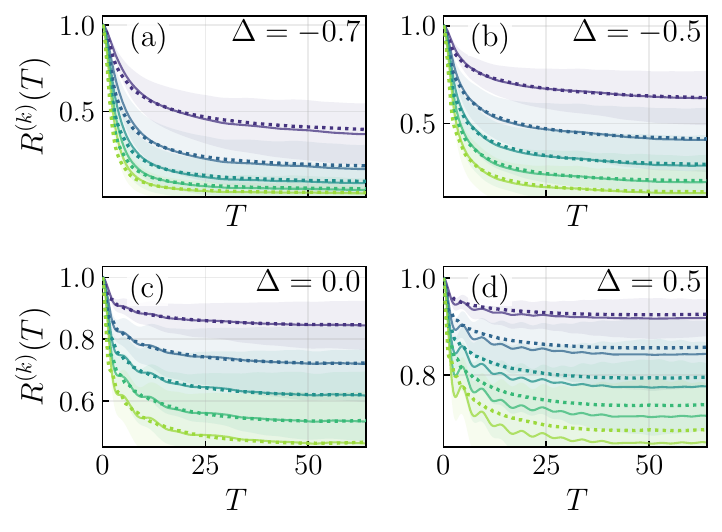}
\caption{
Time dependence of the normalized FP
$R^{(k)}(T)$ for the interacting XXZ chain with $L=64$, $J=1$ and $\sigma_h=0.05$ at
$\Delta=-0.7,-0.5,0.0,0.5$ with panels (a)–(d). For each panel from top to bottom, the different colors label $k=1 \dots 5$.
Solid curves with shaded uncertainty are TEBD numerics. Dotted curves are the TLL prediction from Eq.~\eqref{eq:main_result}.
For each $\Delta$, a single UV cutoff $\alpha$ is obtained from a $k=1$ fit in an intermediate-time window and then kept fixed for all $k$.
}
\label{fig:XXZ}
\end{figure}

We now turn to the interacting chain at several values of $\Delta$. Figure~\ref{fig:XXZ} shows the FP ratio for four representative anisotropies.
The trend across panels reflects the monotonic dependence of $g$ on $\Delta$ within the gapless regime $|\Delta|<1$. As $\Delta$ decreases from $0.5$ toward $-1$, $K$ grows and $u$ shrinks, so $g$ increases and the FP decays faster, which confirms that the strongest randomness is realized near Heisenberg ferromagnetic point $\Delta\to -1$.
For $\Delta=-0.7$ and $-0.5$ in Fig.~\ref{fig:XXZ}(a,b), where $K>3/2$ and the backscattering term is irrelevant~\cite{Giamarchi2004, GiamarchiSchulz1988}, the field-theory curves closely track the TEBD data for all $k=1,\dots,5$. At $\Delta=0$ in Fig.~\ref{fig:XXZ}(c), the agreement remains quantitative, consistent with the free-fermion benchmark. For $\Delta=0.5$ in Fig.~\ref{fig:XXZ}(d), $K=3/4$ and $g$ is smallest, so the overall decay is weakest. However, in this case, the TLL curves show systematic deviation from the TEBD numerics, signaling that the forward-scattering field theory cannot consistently describe the effects at this anisotropy. Significant oscillations are also visible, which we attribute to backward scattering becoming relevant for $K<3/2$. We have verified that employing higher-order filters in the random transversal field further suppresses the backward-scattering component and extends the regime of quantitative agreement to larger $\Delta$. Details are given in the supplemental material~\cite{SM}. Notably, in all panels the single fitted $\alpha$ from $k=1$ correctly predicts the hierarchy of higher-$k$ curves without further adjustment, confirming the $k$-linear structure of Eq.~\eqref{eq:main_result}.


\begin{figure}[t]
\centering
\includegraphics[width=0.75\linewidth]{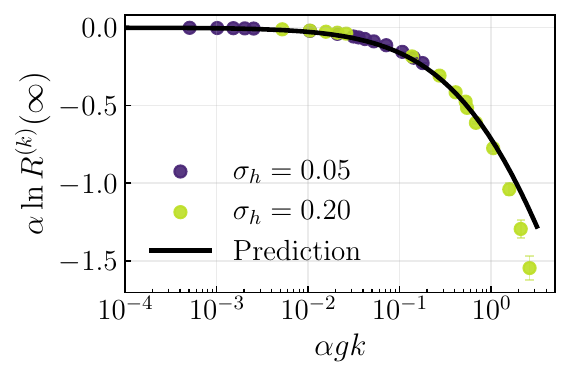}
\caption{
Late-time one-parameter collapse.
From the late-time plateau with last $10\%$ of time points near $T\!\sim\!64$, we plot
$\alpha \ln R^{(k)}(\infty)$ versus $\alpha g k$ with
$g=8\pi K\gamma /u^2$ and $\gamma=\sigma_h^2/\pi^2$,
for $k=1,\dots,5$, $\sigma_h=0.05,0.2$, and selected anisotropic parameters $\Delta=-0.7,0.0,0.5$.
The black curve is the analytic prediction from
Eq.~\eqref{eq:Rinf_closed}, with no additional fitting.
}
\label{fig:plateau}
\end{figure}

As seen in Fig.~\ref{fig:XXZ}, the FP ratio decays to a saturated value, consistent with the predicted late-time plateau. For a finite system of size $L$, the late-time FP ratio $R^{(k)}(\infty)$ should be understood as $R^{(k)}(L)$, while the well-defined $R^{(k)}(\infty)$ of Eq.~\eqref{eq:Rinf_closed} is obtained in the thermodynamic limit $L\to \infty$. We extract the numerical plateau by averaging the last $10\%$ of time points near $T\sim 64$. As shown in Fig.~\ref{fig:plateau}, plotting $\alpha \ln R^{(k)}(\infty)$ versus $\alpha g k$ for $k=1,\dots,5$, $\sigma_h=0.05,\,0.2$, and $\Delta=-0.7,\,0.0,\,0.5$ produces a one-parameter data collapse well captured by Eq.~\eqref{eq:Rinf_closed}. The deviation at large gg
g arises because the FP ratio approaches machine precision, causing the numerical estimate of the plateau to become dominated by statistical fluctuations in the disorder average, as verified in the SM~\cite{SM}.

\begin{figure}
    \centering
    \includegraphics[width=1\linewidth]{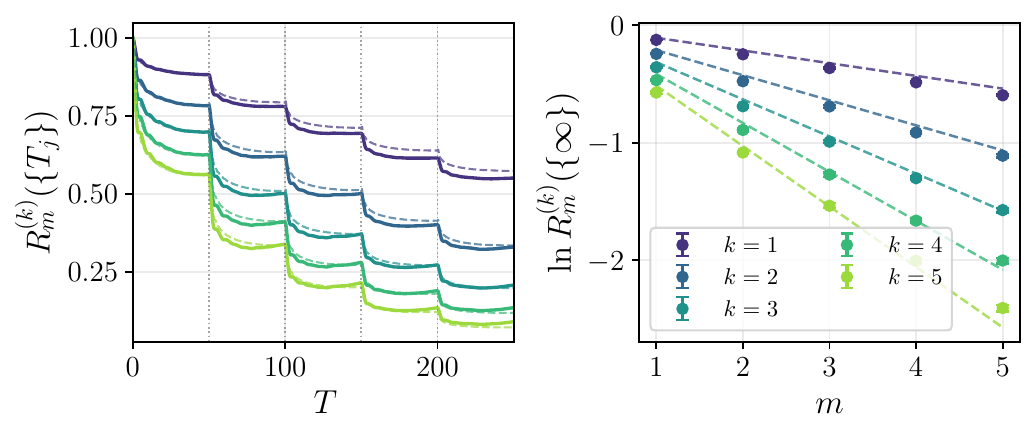}
    \caption{Multiple-quench FP decay with $\Delta=0$, $L=50$, $\sigma_h=0.05$, and $p=1$.
(a)~Normalized FP ratio $R_m^{(k)}(\{T_j\})$ versus total evolution time $T=\sum_j T_j$ for four successive quenches, with vertical gridlines marking the quench times. Different colors label $k=1,\dots,5$. Solid lines are numerics and dashed lines are the field-theory prediction from Eq.~\eqref{eq:main_result_mq}.
(b)~Late-time plateau $\ln R_m^{(k)}(\{\infty\})$ versus number of quenches $m$ for $k=1,\dots,5$. Dots are numerical data and dashed lines are the analytic prediction, confirming the linear-in-$m$ scaling.}
    \label{fig:multiquench}
\end{figure}

We also verify the multiple-quench formula using the free-fermion case $\Delta=0$, which permits long-time simulations. As shown in Fig.~\ref{fig:multiquench}(a), the FP ratio drops to a new plateau after each quench, and the field-theory prediction agrees well with the numerics for all $k=1,\dots,5$ using a single fitted $\alpha$. The late-time plateau values in Fig.~\ref{fig:multiquench}(b) exhibit the linear-in-$m$ scaling in the log plot. The numerics exactly agree with the theoretical prediction, proving that multiple quenches exponentially improve the randomness of the unitary evolution.

\emph{\color{blue}Outlook---}
Complementing previous studies of random unitary ensembles based on nonlocal or stochastic models, we have obtained exact, nonperturbative expressions for the FP ratio of a disordered TLL with local couplings, both for a single quench and for a multiple-quench protocol, and verified them against microscopic random XXZ chain numerics. We conclude that maximum randomness is achieved near the Heisenberg ferromagnetic limit.
Several directions remain open.
First, incorporating backward scattering perturbatively could explain the deviations observed at $\Delta=0.5$ and map out the boundary of the forward-scattering regime.
Second, the analytical dependence of the frame potential on the TLL parameters $K$ and $u$, together with the additive structure of the multiple-quench protocol, suggest to design a practical route to optimize $k$-design convergence in cold-atom~\cite{PhysRevLett.120.050406} or superconducting-circuit experiments~\cite{Roushan2017}, where disorder can be engineered through programmable potential landscapes.
Finally, extending the present calculation to finite $\beta$ to connect with the infinite-temperature $k$-design criterion may yield further exactly solvable benchmarks for quantum randomness and sharpen the connection to randomized-measurement protocols~\cite{Vermersch_2018,Vermersch_2019}.

\emph{\color{blue} Acknowledgments---} We thank Yi-Neng Zhou and Julian Sonner for helpful discussions. This work was supported by the Swiss National
Science Foundation under Division II (Grant No. 200020-
219400).

\bibliography{ref}

\clearpage
\onecolumngrid
\appendix

	\section{Keldysh formulation of the frame potential}

We consider a Tomonaga-Luttinger liquid (TLL) with quenched forward-scattering disorder, described by the Hamiltonian
\begin{equation}
	H_X = H_{\text{TLL}}
	+ H_{\text{dis}}^{(X)},
	\qquad X=U,V,
\end{equation}
where the TLL equation is
\begin{equation}\label{eq:ham_LL}
	H_{\text{TLL}} = \frac{u}{2\pi}\int dx\left[ K(\partial_x\theta)^2 + K^{-1}(\partial_x\phi)^2\right],
\end{equation}
and the disorder Hamiltonian is 
\begin{equation}\label{eq:ham_dis_int}
	H_{\text{dis},X} = \int dx\left[\xi_X(x)\partial_x \phi(x) \right]
\end{equation}
and the static random fields $\xi_X(x)$ are independent Gaussian variables with
\begin{equation}
	\overline{\xi_X(x)\xi_Y^*(x')} = \delta_{XY}\,\gamma\,\delta(x-x') ,
\end{equation}
where $\gamma$ is the disorder strength.

\begin{figure}[h]
    \centering
    \includegraphics[width=0.4\linewidth]{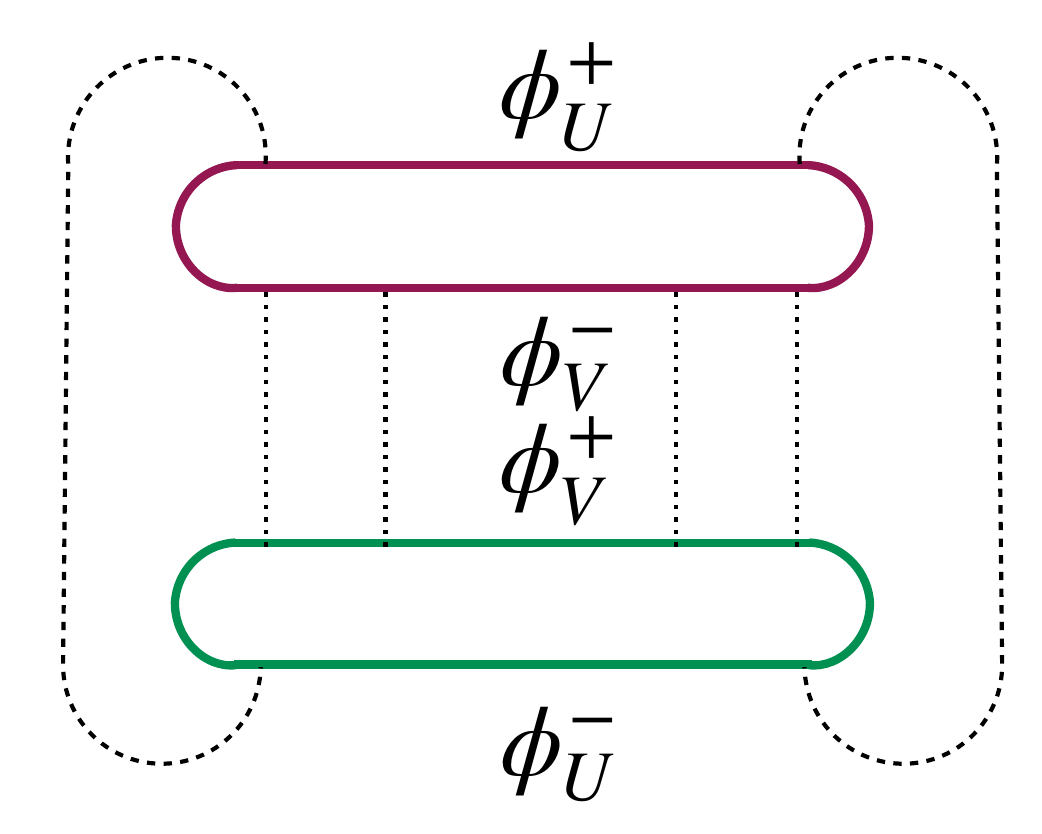}
    \caption{The illustration of the Keldysh contour for the FP with $k=1$. The solid lines represent the contours and the dashed lines represent the disorder coupling.}
    \label{fig_S:contour}
\end{figure}

For fixed disorder realizations, the $k=1$ frame-potential kernel involves
\begin{equation}
	\Tr\!\left(W_U(q)W_V^\dagger(q)\right)
	= \Tr\!\left(e^{-iH_U T} e^{+iH_V T}\right),\qquad
	\Tr\!\left(W_V(q)W_U^\dagger(q)\right)
	= \Tr\!\left(e^{-iH_V T}e^{+iH_U T}\right).
\end{equation}
Each trace is written as a real-time path integral on a closed Keldysh contour. Taken together,
this produces two Keldysh loops, with loop 1 carries $(\phi_U^+(q),\phi_V^-(q))$ and loop2 carries $(\phi_V^+(q),\phi_U^-(q))$.
Here $(+,-)$ denotes the forward and backward branches of the contour, as shown in the Fig.~\ref{fig_S:contour}

We label the four contour fields in the original basis as
\begin{equation}
	\Psi_a(q) =
	\big(\phi_{U}^+(q),\phi_{V}^-(q),
	\phi_{V}^+(q),\phi_{U}^-(q)\big),
	\qquad a=1,\dots,4.
\end{equation}
The corresponding Green's function between loops $\ell,\ell'$ is
\begin{equation}
	G^{(\ell\ell')}(q,q') =
	\begin{pmatrix}
		G^{(\ell\ell')}_{T}(q,q') &
		G^{(\ell\ell')}_{>}(q,q')\\[2pt]
		G^{(\ell\ell')}_{<}(q,q') &
		G^{(\ell\ell')}_{\tilde{T}}(q,q')
	\end{pmatrix}
\end{equation}

On each loop we perform a Keldysh rotation between the pair of fields that are
connected by the trace:
\begin{align}
	\text{loop 1:}\quad
	&\phi_{1,\mathrm{cl}} = \frac{1}{\sqrt{2}}\left(\phi_U^+ + \phi_V^-\right),\qquad
	\phi_{1,\mathrm{q}}  = \frac{1}{\sqrt{2}}\left(\phi_U^+ - \phi_V^-\right),\\
	\text{loop 2:}\quad
	&\phi_{2,\mathrm{cl}} = \frac{1}{\sqrt{2}}\left(\phi_V^+ + \phi_U^-\right),\qquad
	\phi_{2,\mathrm{q}}  = \frac{1}{\sqrt{2}}\left(\phi_V^+ - \phi_U^-\right).
\end{align}
We collect the four loop-wise Keldysh fields into
\begin{equation}
	\Phi_a^T(q) = 
	\big(\phi_{1,\mathrm{cl}}(q),\phi_{1,\mathrm{q}}(q),
	\phi_{2,\mathrm{cl}}(q),\phi_{2,\mathrm{q}}(q)\big)_a,
\end{equation}
with $a=1,\dots,4$.
After integrating out $\theta$, the clean Luttinger part becomes quadratic in Keldysh form,
\begin{equation}
	S_{\text{TLL}} = \frac{1}{2}\sum_{\omega,q}
	\Phi^T(q,\omega)\,
	\bm{G}_0^{-1}(q,\omega)\,
	\Phi(-q,-\omega),
\end{equation}
with the Green's function
\begin{equation}
    \bm{G}_0^{-1}=\begin{pmatrix}
		0                 & (G_0^A)^{-1} & 0                 & 0\\
		(G_0^R)^{-1}      & (G_0^K)^{-1} & 0                 & 0\\
		0                 & 0            & 0                 & (G_0^A)^{-1}\\
		0                 & 0            & (G_0^R)^{-1}      & (G_0^K)^{-1}
	\end{pmatrix}
\end{equation}
where the retarded, advanced and the Kelydsh component is
\begin{equation}
	(G_0^R)^{-1}(\omega,q)
	= \frac{1}{\pi K}\left(\frac{(\omega+i0)^2}{u} - uq^2\right),\qquad
	G_0^A = (G_0^R)^*,\qquad G_0^K = \coth\left(\frac{\beta\omega}{2}\right) (G_0^R - G_0^A).
    \label{eq:TLL_GF}
\end{equation}
where $\beta$ is the inverse temperature.

Since the action is Gaussian, the correlators are fully encoded in the $4\times4$ Keldysh Green’s
function
\begin{equation}
	G_{ab}(t-t') = -i\,\big\langle \Phi_a(q)\,\Phi_b(q')\big\rangle_0,
	\qquad a,b=1,\dots,4,
\end{equation}
with the usual cl/q structure on each pair of loops. It is convenient to
organize these in terms of $2\times2$ Keldysh blocks between loops
$\ell,\ell'=1,2$,
\begin{equation}
	G_{(\ell\ell')}(q,q') =
	\begin{pmatrix}
		G_{(\ell\ell')}^{\mathrm{cl},\mathrm{cl}}(q,q') &
		G_{(\ell\ell')}^{\mathrm{cl},\mathrm{q}}(q,q')\\[2pt]
		G_{(\ell\ell')}^{\mathrm{q},\mathrm{cl}}(q,q') &
		G_{(\ell\ell')}^{\mathrm{q},\mathrm{q}}(q,q')
	\end{pmatrix}
	=
	\begin{pmatrix}
		G^{K}_{\ell\ell'}(q,q') & G^{R}_{\ell\ell'}(q,q')\\
		G^{A}_{\ell\ell'}(q,q') & 0
	\end{pmatrix}.
\end{equation}
Here $G^{R/A/K}_{\ell\ell'}$ are the retarded, advanced and Keldysh components
between loop $\ell$ and loop $\ell'$.

The forward-scattering disorder term for a given sector $X\in\{U,V\}$ is
\begin{equation}
	H_{X,\text{dis}} = \int dx\,
	\big[\xi_X(x)\partial_x \phi(x)\big],
\end{equation}
with Gaussian weight to be $P[\xi_X]\propto \exp\!\left[-\frac{1}{2\gamma}\int dx\,|\xi_X(x)|^2\right]$.

On the Keldysh contour $C_X$ for this sector, we write
\begin{equation}
	iS_{X,\text{dis}}[\phi_X^\pm,\xi_X]
	= -i\int_{C_X} dt\int dx\,
	\big[\xi_X \partial_x \phi(x)\big] = -i\int dx\,
	\big[\xi_X J_X(x)\big]
\end{equation}
where $J_X(x) = \int_0^T dt\,\big[\partial_x\phi_X^+(x,t) - \partial_x\phi_X^-(x,t)\big]$. 
Integrating out $\xi_X$ gives
\begin{equation}
	\int\mathcal{D}\xi_X\mathcal{D}\xi_X^*\,
	e^{-\frac{1}{2\gamma}\int dx|\xi_X|^2 + iS_{X,\text{dis}}}
	\propto
	\exp\!\left[-\gamma\int dx\,|J_X(x)|^2\right],
\end{equation}
so the effective disorder action for sector $X$ is $S_{X,\text{dis}}^{\text{eff}} = \iu \gamma\int dx\,|J_X(x)|^2$. Expanding the $J_X(x)$ leads to the result
\begin{equation}
	|J_X(x)|^2
	= \int_0^T\!dt\int_0^T\!dt'\,
	\big[\partial_x\phi_X^+(x,t) - \partial_x\phi_X^-(x,t)\big]
	\big[\partial_x\phi_X^+(x,t') - \partial_x\phi_X^-(x,t')\big],
\end{equation}
and using the symmetry $t\leftrightarrow t'$, one can rewrite the four terms
exactly as bilinears of the Keldysh fields on both loops. Passing to momentum space and dropping the position label, the disorder-averaged part of the action reads
\begin{equation}
	\begin{split}
		S^{\text{eff}}_{\text{dis}} &=  \iu \gamma\int \frac{\diff q}{2\pi} \int_0^T\!dt\int_0^T\!dt'\,
		q^2 \Psi^T(q,t)
		\tilde{C}_0 \Psi(-q,t')
		 \\
	\end{split},
\end{equation}
and after the Keldysh rotation
\begin{equation}
	\begin{split}
		S^{\text{eff}}_{\text{dis}} &=  \iu \gamma\int \frac{\diff q}{2\pi} \int_0^T\!dt\int_0^T\!dt'\,
		q^2 \Phi^T(q,t)
		C_0 \Phi(-q,t')
		\\
	\end{split},
\end{equation}
with the bare coupling matrices in the original and Keldysh bases given by
\begin{equation}
    \tilde{C}_0 = \begin{pmatrix}
			1 & 0 & 0 & -1 \\
			0 & 1 & -1 & 0 \\
			0 & -1  & 1 & 0 \\
			-1 & 0 & 0 & 1 \\
		\end{pmatrix},\qquad 
    C_0 = \begin{pmatrix}
			1 & 0 & -1 & 0 \\
			0 & 1 & 0 & 1 \\
			-1 & 0 & 1 & 0 \\
			0 & 1 & 0 & 1 \\
		\end{pmatrix}   
\end{equation}

After Fourier transformation to frequency space,
\begin{equation}
	\begin{split}
		S^{\text{eff}}_{\text{dis}} &=  \iu \gamma\int \frac{\diff q}{2\pi}\int \frac{\diff \omega}{2\pi} \int \frac{\diff \omega'}{2\pi}
		\frac{q^2\left(1-e^{\iu \omega T}\right)\left(1-e^{-\iu \omega' T}\right)}{\omega \omega'} \Phi^T(q,\omega)
		C_0 \Phi(-q,-\omega')
		\\
	\end{split}
\end{equation}

Therefore the total action is
\begin{equation}
	\begin{split}
		S = S_{\text{TLL}} + S_{\text{dis}} =& \frac{1}{2}\sum_{\omega,q}
		\Phi^T(q,\omega)\,
		\bm{G}_0^{-1}(q,\omega)\,
		\Phi(-q,-\omega) + \\
		 &  \iu \gamma\int \frac{\diff q}{2\pi}\int \frac{\diff \omega}{2\pi} \int \frac{\diff \omega'}{2\pi}
		\frac{q^2\left(1-e^{\iu \omega T}\right)\left(1-e^{\iu \omega' T}\right)}{\iu^2\omega \omega'} \Phi^T(q,\omega)
		C_0 \Phi(-q,\omega')
		\\
	\end{split}
\end{equation}

Starting from the quadratic action (for each momentum $q$),
\begin{equation}
	S = \frac{1}{2}\int\!\frac{d\omega}{2\pi}\frac{d\omega'}{2\pi}\,
	\Phi^T(q,\omega)\,\bm{\Omega}(q;\omega,\omega';T)\,\Phi(-q,\omega'),
\end{equation}
we separate the kernel into the Luttinger-liquid part and the disorder part,
\begin{equation}
	\bm{\Omega}(q;\omega,\omega';T)
	= \bm{\Omega}_0^{-1}(q;\omega,\omega')
	+ \Sigma(q;\omega,\omega';T),
\end{equation}
where $\bm{\Omega}_0^{-1}$ is the block-diagonal Keldysh kernel in frequency obtained from $S_{\text{TLL}}$,
\begin{equation}
	\bm{\Omega}_0^{-1}(q;\omega,\omega') = (2\pi)\,\delta(\omega+\omega')\,\bm{G}_0^{-1}(q,\omega).
\end{equation}
Its inverse is
 \begin{equation}
 	\bm{\Omega}_0(q;\omega,\omega') = (2\pi)\,\delta(\omega+\omega')\,\bm{G}_0(q,\omega),
 \end{equation}
 where
 \begin{equation}
 	\bm{G}_0(q,\omega) =
 	\begin{pmatrix}
 		G_0^K           & G_0^R & 0                 & 0\\
 		G_0^A            &  0 & 0                 & 0\\
 		0                 & 0            & 	G_0^K           & G_0^R\\
 		0                 & 0            & 	G_0^A            &  0
 	\end{pmatrix}.
 \end{equation}

 The frequency off-diagonal disorder term is
\begin{equation}
	\Sigma(q;\omega,\omega';T)
	= g_T(\omega)\,\bm{C}_0(q) \,g_T(\omega')
\end{equation}
with $g_T(\omega) = \frac{1 - e^{i\omega T}}{\omega}$ and
\begin{equation}
	\bm{C}_0(q) = -\iu \gamma q^2
	\begin{pmatrix}
		1 & 0 & -1 & 0 \\
		0 & 1 & 0 & 1 \\
		-1 & 0 & 1 & 0 \\
		0 & 1 & 0 & 1
	\end{pmatrix}.
\end{equation}
The overall prefactor is fixed by matching to $S_{\text{dis}}$. The manipulations below only use the separable structure in frequency.

The disorder-averaged first frame potential is represented by the Gaussian functional integral
\begin{equation}
	F^{(1)}(T)
	= \prod_q \int\!\mathcal{D}\Phi(q)\,
	\exp\bigg\{\frac{i}{2}\int\!\frac{d\omega}{2\pi}\frac{d\omega'}{2\pi}\,
	\Phi^T(q,\omega)\,\bm{\Omega}(q;\omega,\omega';T)\,\Phi(-q,-\omega')\bigg\}.
\end{equation}
Since the action is quadratic, the functional integral is proportional to the inverse square root of the determinant of the kernel,
\begin{equation}
	F^{(1)}(T) \propto \prod_q \big[\det \bm{\Omega}(q;T)\big]^{-1/2}.
\end{equation}

The separable frequency dependence of the disorder kernel can be written as a rank-one update by defining $|g_T(\omega)\rangle$ as the vector with components $g_T(\omega)$ in frequency space, and setting $\mathcal{U}=|g_T(\omega)\rangle\otimes\id_4$ and $\mathcal{V}=\langle g_T(\omega')|\otimes\id_4$. We then exploit the rank-one structure $\bm{\Omega} = \bm{\Omega}_0^{-1} + \mathcal{U}\, \bm{C}_0\, \mathcal{V}$ and the matrix determinant lemma. Using $\det(\id_m + A B) = \det(\id_n + B A)$ for $A\in \mathbb{C}^{m\times n}$ and $B \in \mathbb{C}^{n\times m}$, we identify $A = \bm{\Omega}_0 \mathcal{U}$ and $B = \bm{C}_0 \mathcal{V}$ and obtain
\begin{equation}
	\det\bm{\Omega}
	= \det \bm{\Omega}_0^{-1}\,
	\det\big(\id_{4N_{\omega}} + \bm{\Omega}_0 \mathcal{U}\, \bm{C}_0\, \mathcal{V}\big)
	= \det \bm{\Omega}_0^{-1}\,
	\det\big(\id_4 + \bm{C}_0\, \mathcal{V}\, \bm{\Omega}_0\, \mathcal{U}\big).
\end{equation}
The $4\times4$ polarization matrix in Keldysh$\times$loop space is
\begin{equation}
	\bm{\Pi}(q;T)
	\equiv \mathcal{V}\, \bm{\Omega}_0\, \mathcal{U}
	= \int\!\frac{d\omega}{2\pi}\,
	g_T(-\omega)\,\bm{G}_0(q,\omega)\,g_T(\omega),
\end{equation}
which can be evaluated directly thanks to the block-diagonal structure of $\bm{\Omega}_0$.
We thus find
\begin{equation}
	\det \bm{\Omega}(q;T)
	= \det \bm{\Omega}_0^{-1}(q)\,
	\det\big(\id_4 + \bm{C}_0(q)\,\bm{\Pi}(q;T)\big).
\end{equation}
The factor $\det \bm{\Omega}_0^{-1}$ is independent of $T$. Normalizing the frame potential by its value at $T=0$, where $g_T(\omega)=0$ and hence $\bm{\Pi}(q;0)=0$, we obtain
\begin{equation}
	R^{(1)}(T) \equiv \frac{F^{(1)}(T)}{F^{(1)}(0)}
	= \prod_q
	\left(
	\det\big(\id_4 + \bm{C}_0(q)\,\bm{\Pi}(q;T)\big)
	\right)^{-1/2}.
\end{equation}
Equivalently,
\begin{equation}
	\label{eq:logF_final}
	\ln R^{(1)}(T)
	= -\frac{1}{2}\sum_q
	\ln\det\big[\id_4 + \bm{C}_0(q)\,\bm{\Pi}(q;T)\big].
\end{equation}

Equation~\eqref{eq:logF_final} shows explicitly that the time dependence of the frame potential is controlled solely by the separable disorder contribution via the $4\times4$ matrix $\bm{C}_0(q)\,\bm{\Pi}(q;T)$, while the clean kernel $\bm{\Omega}_0$ only contributes an overall $T$-independent normalization.

\section{Evaluation of FP in the TLL}

The product $\bm{C}_0(q)\,\bm{\Pi}(q;T)$ reads
\begin{equation}
	\bm{C}_0(q)\,\bm{\Pi}(q;T) = -\iu \gamma q^2 \int \frac{\diff\omega}{2\pi}
	\begin{pmatrix}
		1 & 0 & -1 & 0 \\
		0 & 1 & 0 & 1 \\
		-1 & 0 & 1 & 0 \\
		0 & 1 & 0 & 1
	\end{pmatrix}
	\begin{pmatrix}
		G_0^K           & G_0^R & 0                 & 0\\
		G_0^A            &  0 & 0                 & 0\\
		0                 & 0            & 	G_0^K           & G_0^R\\
		0                 & 0            & 	G_0^A            &  0
	\end{pmatrix} g_T(-\omega)g_T(\omega).
\end{equation}

Carrying out the matrix multiplication and evaluating the $4\times 4$ determinant, only the Keldysh component $G_0^K$ survives, giving the frame-potential ratio in the compact form
\begin{equation}
	\label{eq:logF_reduced_LL}
	\ln R^{(1)}(T)
	= -\frac{1}{2}\sum_{q>0}
	\ln\Big[1 - 2 \iu \gamma q^2 \int\!\frac{d\omega}{2\pi}\,g_T(\omega)g_T(-\omega)\,G_0^K(q,\omega)\Big].
\end{equation}

For the Luttinger liquid, based on Eq.~\eqref{eq:TLL_GF}, we can derive the Keldysh Green's function. We introduce the mode frequency $E_q \equiv u|q|$.
From $G_0^R$ one finds the spectral identity
\begin{equation}
	G_0^R(\omega,q) - G_0^A(\omega,q)
	= \iu\,\frac{\pi^2 K}{|q|}
	\big[\delta(\omega - E_q) - \delta(\omega + E_q)\big],
\end{equation}
so that
\begin{equation}
	G_0^K(\omega,q)
	= \iu\,\frac{\pi^2 K}{|q|}
	\coth\left(\frac{\beta\omega}{2}\right)
	\big[\delta(\omega - E_q) - \delta(\omega + E_q)\big].
\end{equation}

For the single-quench time window $g_T(\omega)$ defined above, the product simplifies to
\begin{equation}
    g_T(\omega)g_T(-\omega)
	= \frac{4\sin^2(\omega T/2)}{\omega^2}.
\end{equation}

Substituting the spectral representation of $G_0^K$ and performing the frequency integral via the $\delta$-functions, we arrive at
\begin{equation}
	\det\big[\id_4 + \bm{C}_0(q)\,\bm{\Pi}(q;T)\big]
	= 1 - 2 \iu\, \gamma q^2 \left[
	\iu\,\frac{4\pi K}{|q|}\,
	\frac{\sin^2\big(u|q| T/2\big)}{(u|q|)^2}
	\coth\left(\frac{\beta u|q|}{2}\right)
	\right].
\end{equation}
This simplifies to
\begin{equation}
	\det\big[\id_4 + \bm{C}_0(q)\,\bm{\Pi}(q;T)\big]
	= 1 + \frac{8\pi K \gamma}{u^2 |q|}\,
	\sin^2\left(\frac{u|q| T}{2}\right)
	\coth\left(\frac{\beta u|q|}{2}\right).
\end{equation}
Hence the frame-potential ratio is
\begin{equation}
		\ln R^{(1)}(T)
		= -\frac{1}{2}\sum_{q>0}
		\ln\left[
		1 + A_q(T)
		\right] e^{-\alpha|q|}
\end{equation}
with
\begin{equation}\label{eq:AqT_SM}
	A_q(T)
	=
	\frac{g}{|q|}
	\sin^2\left(\frac{u|q| T}{2}\right)
	\coth\left(\frac{\beta u|q|}{2}\right),
\end{equation}
where $g\equiv 8\pi K\gamma/u^2$ is the dimensionless coupling defined in the main text, and $\alpha$ is the ultraviolet cutoff in momentum space.

Below we discuss the behavior in the limits of small and large disorder strength $\gamma$, and then the long time limit.

\subsection{Short time limit} One has $A_q(T)\ll 1$ for all relevant $q$ and $T$, so one may expand the logarithm,
\begin{equation}
	\ln(1+A_q) = A_q  + \mathcal{O}(A_q^2).
\end{equation}
To leading order in $\gamma$,
\begin{equation}
	\ln R^{(1)}(T)
	\simeq
	-\frac{1}{2}\sum_{q>0} A_q(T)\,e^{-\alpha|q|}
	+ \mathcal{O}(\gamma^2),
\end{equation}
which, upon substituting the explicit form of $A_q$, reads
\begin{equation}
	\ln R^{(1)}(T)
	\simeq
	-\frac{g}{2}\sum_{q>0}
	\frac{1}{|q|}
	\sin^2\left(\frac{u|q| T}{2}\right)
	\coth\left(\frac{\beta u|q|}{2}\right) e^{-\alpha|q|}.
\end{equation}
Thus at weak disorder the frame potential decays perturbatively,
\emph{linearly} in $\gamma$.

In the low-temperature limit, $\coth(\beta u |q|/2)\to 1$. Passing to the thermodynamic limit $\sum_{q>0}\to\frac{L}{\pi}\int_0^\infty dq$, the regulated integral gives
\begin{equation}
	\frac{gL}{2}\int_{0}^{\infty}\frac{\diff q}{\pi} e^{-q\alpha}
	\frac{1}{q}
	\sin^2\left(\frac{uq T}{2}\right)
	= \frac{L\gamma K}{u^2}\log\left( \frac{\alpha^2 +u^2 T^2}{\alpha^2} \right),
\end{equation}
which yields
\begin{equation}
	R^{(1)}(T) = \left( \frac{\alpha^2 +u^2 T^2}{\alpha^2} \right)^{-L\gamma K/u^2}.
\end{equation}


\subsection{Long time limit}
For $uT\gg\alpha$ the factor $\sin^2(uqT/2)$ oscillates rapidly and self-averages in the momentum integral. Passing to the thermodynamic limit $\sum_{q>0}\to\frac{L}{\pi}\int_0^\infty dq$ and taking the low-temperature limit $\coth(\beta u|q|/2)\to 1$, the frame-potential ratio becomes
\begin{equation}
	\ln R^{(1)}(T)
	= -\frac{L}{2\pi} \int_0^{\infty} \diff{q}\, e^{-\alpha q}
	\ln\left[
	1 + \frac{g}{q}
	\sin^2\!\left(\frac{uq T}{2}\right) \right].
\end{equation}
In the late-time plateau we take $T\to\infty$ and replace $\sin^2$ by its phase average. Setting $\theta=uqT/2$ as a uniformly distributed phase gives the angular integral identity
\begin{equation}
	\frac{1}{2\pi}\int_0^{2\pi} \diff{\theta}\, \ln\!\left(1 + \frac{g}{q}\sin^2\theta \right) = 2\ln \left( \frac{1+ \sqrt{1+g/q}}{2} \right).
\end{equation}
The late-time plateau therefore reads
\begin{equation}
	\ln R^{(1)}(\infty) =
	-\frac{L}{\pi} \int_0^{\infty} \diff{q}\, e^{-\alpha q}
	\ln \left( \frac{1+ \sqrt{1+g/q}}{2} \right).
\end{equation}
To evaluate this integral analytically, we differentiate with respect to $g$,
\begin{equation}
	\frac{\partial}{\partial g}\ln R^{(1)}(\infty) = - \frac{L}{2\pi} \int_0^{\infty} \diff{q}\, e^{-\alpha q} \frac{1}{g}\left(1 - \sqrt{\frac{q}{g+q}} \right) =- \frac{L}{2\pi} \frac{1}{g\alpha}\left(1-\frac{\sqrt{\pi}\, U\!\left(\tfrac{1}{2},0,g\alpha \right)}{2 } \right),
\end{equation}
where $U$ is the confluent hypergeometric function of the second kind.
Integrating back with the boundary condition $\ln R^{(1)}(\infty)\to 0$ as $g\to 0$ yields
\begin{equation}
	\ln R^{(1)}(\infty) = -\frac{L}{2\pi\alpha}\left( \ln\!\left(\frac{\alpha g}{4}\right)+e^{\alpha g/2} K_0\!\left(\frac{\alpha g}{2}\right)+\gamma_E \right),
\end{equation}
where $K_0$ is the modified Bessel function of the second kind and $\gamma_E$ is the Euler-Mascheroni constant. The generalization to arbitrary $k$ is obtained by replacing $g\to gk$ and is derived in multiple-quenched section below.

\subsection{Strong disorder $g\gg 1$.} One has $A_q(T)\gg 1$ for most contributing modes (except at very short times or very large $|q|$), so
\begin{equation}
	\ln\left[1+A_q(T)\right]
	\simeq
	\ln A_q(T).
\end{equation}
Then
\begin{equation}
	\ln R^{(1)}(T)
	\simeq
	-\frac{1}{2}\sum_{q>0} \ln A_q(T)\,e^{-\alpha|q|},
\end{equation}
which, upon expanding $\ln A_q$, becomes
\begin{equation}
	\ln R^{(1)}(T)
	\simeq
	-\frac{1}{2}\sum_{q>0}
	\left[
	\ln\left(\frac{g}{|q|} \coth\left(\frac{\beta u|q|}{2}\right)\right)
	+
	\ln\sin^2\left(\frac{u|q|T}{2}\right)
	\right]e^{-\alpha|q|}.
\end{equation}
In particular, the $\gamma$-dependence is
\begin{equation}
	\ln R^{(1)}(T)
	\sim
	-\frac{N_q}{2}\ln g + \dots,
\end{equation}
where $N_q$ is the number of $q$ modes effectively contributing which is set by the UV cutoff $\alpha$. Thus at strong disorder the frame potential is suppressed very strongly. In a continuum limit this drives $R^{(1)}(T)$ to zero exponentially in system size, corresponding to very strong scrambling.

\section{Generalization to arbitrary $k$}

We now generalize the derivation to the $k$th frame potential $F^{(k)}(T) = \mathbb{E}|\mathcal{Z}(T)|^{2k}$, which involves $2k$ Keldysh contours.
The disorder-averaged effective action on the $4k$-component field acquires the structure
\begin{equation}
	\begin{split}
	    \iu S_{\text{dis}}
	= -\gamma \int \diff{x} \int_0^T\!dt\int_0^T\!dt'\,
	&\Bigg( \left[\sum_{i=0}^{k-1} \partial_x\Psi_{4i+1}(x,t) - \partial_x\Psi_{4i+4}(x,t)\right]
	\left[\sum_{i=0}^{k-1} \partial_x\Psi_{4i+1}(x,t) - \partial_x\Psi_{4i+4}(x,t)\right] \\
    &-  \left[\sum_{i=0}^{k-1} \partial_x\Psi_{4i+2}(x,t) - \partial_x\Psi_{4i+3}(x,t)\right]
	\left[\sum_{i=0}^{k-1} \partial_x\Psi_{4i+2}(x,t) - \partial_x\Psi_{4i+3}(x,t)\right]  \Bigg)
	\end{split}
\end{equation}

Here and below, $\operatorname{Diag}(A_1,\dots,A_n)$ denotes the block-diagonal matrix with diagonal blocks $A_1,\dots,A_n$.
The Keldysh rotation on each replica block is encoded in the $4\times4$ matrix $M = \operatorname{Diag}(m,m)$ with $m = \frac{1}{\sqrt{2}}\begin{pmatrix} 1 & 1 \\ 1 & -1 \end{pmatrix}$, so that $\Phi = M^{-1}\Psi$ on each replica.
The clean and disorder parts of the $4k\times 4k$ kernel generalize to
\begin{equation}
    \bm{\Omega}_0^{(k),-1}(q;\omega,\omega') = (2\pi)\delta(\omega+\omega') \operatorname{Diag}(\underbrace{\bm{G}_0(q,\omega), \cdots, \bm{G}_0(q,\omega)}_{k})
\end{equation}
and
\begin{equation}
    \bm{C}_0^{(k)} = \underbrace{\operatorname{Diag}(M, \cdots, M)}_k
    \underbrace{\begin{pmatrix}
        \tilde{C}_0 & \tilde{C}_0 & \cdots & \tilde{C}_0 \\ 
        \tilde{C}_0 & \ddots & \cdots & \tilde{C}_0 \\
        \tilde{C}_0 & \tilde{C}_0 & \cdots & \tilde{C}_0 \\
    \end{pmatrix}}_{k}
    \underbrace{\operatorname{Diag}(M, \cdots, M)}_k
\end{equation}

Applying the same determinant-lemma reduction as for $k=1$, one finds that the $4k\times4k$ determinant factorizes into a product of $4\times4$ blocks, each contributing $\ln(1+kA_q)$ rather than $\ln(1+A_q)$. The normalized FP ratio therefore becomes
\begin{equation}
	\ln R^{(k)}(T)
		= -\frac{1}{2}\sum_{q>0}
		\ln\left[
		1 + k A_q(T)
		\right] e^{-\alpha|q|}
\end{equation}
with $A_q(T)$ defined in Eq.~\eqref{eq:AqT_SM}, reproducing Eq.~\eqref{eq:main_result} of the main text.
Consequnetly, replacing $g$ with $g k$ reproducing all the $k$-th FP result in the main text.

\section{Multiple-quench protocol}
\label{sec:multiquench}

We now extend the single-quench result to the multiple-quench protocol defined in the main text. The total evolution consists of $m$ successive segments, each governed by an independently drawn disorder realization. For the $j$th segment, the time-window function becomes
\begin{equation}
g_j(\omega)
=
e^{i\omega\tau_{j-1}}
\frac{1-e^{i\omega T_j}}{\omega}
\label{eq:gj_def}
\end{equation}
with $\tau_j\equiv \sum_{\ell=1}^j T_\ell$ and 
$\tau_0\equiv 0$
After averaging over disorder, the kernel for each momentum mode takes the separable form
\begin{equation}
\bm{\Omega}(q)
=
\bm{\Omega}_0^{-1}(q)
+
\sum_{j=1}^m
\mathcal{U}_j\,\bm{C}_0(q)\,\mathcal{V}_j
\label{eq:Omega_mq}
\end{equation}
with $\mathcal{U}_j=|g_j\rangle\otimes\id_4$ and $
\mathcal{V}_j=\langle g_j|\otimes\id_4$
Because different quenches are independent, the disorder correlator is diagonal in the quench index,
\begin{equation}
\overline{\xi_{X,j}(x)\xi_{Y,\ell}(x')}
=
\gamma\,\delta_{XY}\delta_{j\ell}\delta(x-x'),
\end{equation}
so there are no cross terms between different segments.
Applying the matrix determinant lemma then gives
\begin{equation}
\det \bm{\Omega}(q)
=
\det \bm{\Omega}_0^{-1}(q)
\prod_{j=1}^m
\det\!\left(\id_4+\bm{C}_0(q)\,\bm{\Pi}_j(q)\right),
\label{eq:det_mq}
\end{equation}
with
\begin{equation}
\bm{\Pi}_j(q)
=
\int \frac{\diff\omega}{2\pi}\,
g_j(-\omega)\,\bm{G}_0(q,\omega)\,g_j(\omega).
\label{eq:Pi_j_def}
\end{equation}
The phase factor in Eq.~\eqref{eq:gj_def} drops out from $g_j(-\omega)g_j(\omega)$, so that
\begin{equation}
\bm{\Pi}_j(q)
=
\int \frac{\diff\omega}{2\pi}\,
g_{T_j}(-\omega)\,\bm{G}_0(q,\omega)\,g_{T_j}(\omega),
\end{equation}
namely each block is identical to the single-quench function with $T\to T_j$.
Therefore,
\begin{equation}
\ln \det\!\left(\id_4+\bm{C}_0(q)\,\bm{\Pi}_j\right)
=
\ln\!\bigl[1+A_q(T_j)\bigr],
\label{eq:block_Aq}
\end{equation}
where $A_q(T)$ is exactly the single-quench function in Eq.~\eqref{eq:AqT_SM}.

As a result, the normalized ratio factorizes over quench segments:
\begin{equation}
\ln R_m^{(1)}(\{T_j\})
=
-\frac{1}{2}\sum_q\sum_{j=1}^m
\ln\!\bigl[1+A_q(T_j)\bigr]\,e^{-\alpha|q|}.
\label{eq:R1_mq}
\end{equation}
Replicating the contour to $2k$ loops gives the general result
\begin{equation}
\ln R_m^{(k)}(\{T_j\})
=
-\frac{1}{2}\sum_q\sum_{j=1}^m
\ln\!\bigl[1+kA_q(T_j)\bigr]\,e^{-\alpha|q|},
\label{eq:Rk_mq}
\end{equation}
which is Eq.~\eqref{eq:main_result_mq} of the main text.

\section{Higher-order filter construction}

In the main text, the forward-scattering condition is enforced at filter order $p=1$ by convolving the i.i.d.\ Gaussian disorder $h_i\sim\mathcal{N}(0,\sigma_h^2)$ with the two-point kernel $f_n^{(1)}=\delta_{n,0}+\delta_{n,1}$, yielding $\tilde{h}_i^{(1)}=(h_{i-1}+h_i)/2$ and a power spectrum $S_1(q)=\cos^2(q/2)\,\sigma_h^2$ that vanishes at $q=\pi$.
Here we describe the generalization to arbitrary order $p$.

We define the order-$p$ filtered field as
\begin{equation}\label{eq:filter_def}
  \tilde{h}_i^{(p)} = \sum_{m=0}^{p} c_m^{(p)}\, h_{i-m},
\end{equation}
where the coefficients follow a normalized binomial kernel,
\begin{equation}\label{eq:filter_coeff}
  c_m^{(p)} = \frac{\binom{p}{m}}{2^p}.
\end{equation}
The normalization $\sum_m c_m^{(p)} = 1$ ensures that $\hat{f}_p(0)=1$ and hence $S_p(0) = \sigma_h^2$ for all $p$, preserving the continuum white-noise identification $\gamma = \sigma_h^2/\pi^2$.
For $p=1$ this reduces to $c_0=c_1=1/2$, recovering the main-text construction.

The Fourier transform of the kernel is
\begin{equation}\label{eq:filter_fourier}
  \hat{f}_p(q) = \sum_{m=0}^{p} c_m^{(p)}\, e^{-\iu mq}
  = \frac{(1+e^{-\iu q})^p}{2^p},
\end{equation}
so the disorder power spectrum of the filtered field reads
\begin{equation}\label{eq:filter_spectrum}
  S_p(q) = |\hat{f}_p(q)|^2\,\sigma_h^2
  = \cos^{2p}(q/2)\,\sigma_h^2.
\end{equation}
Near the zone boundary the spectrum vanishes as $S_p(q)\propto (\pi-q)^{2p}$ for $q\to\pi$, providing a $2p$th-order zero at $q=\pi$.
Since disorder at momentum $q\approx 2k_F=\pi$ drives backward scattering~\cite{GiamarchiSchulz1988, Giamarchi2004}, this stronger suppression with increasing $p$ progressively reduces the backward-scattering amplitude and brings the lattice model closer to the pure forward-scattering limit assumed in the field theory.

For a chain with open boundary conditions of length $L$, the convolution in Eq.~\eqref{eq:filter_def} requires values $h_{i-m}$ with $i-m<0$ near the left boundary, and analogously near the right boundary for more general case. We adopt a reflective  extension, setting $h_{-n}=h_{n}$ for $n>0$, which preserves the local variance and avoids artificial edge effects. The resulting boundary correction is of order $\mathcal{O}(p/L)$ and negligible for the system sizes considered here.

\begin{figure}[t]
\centering
\includegraphics[width=1.0\linewidth]{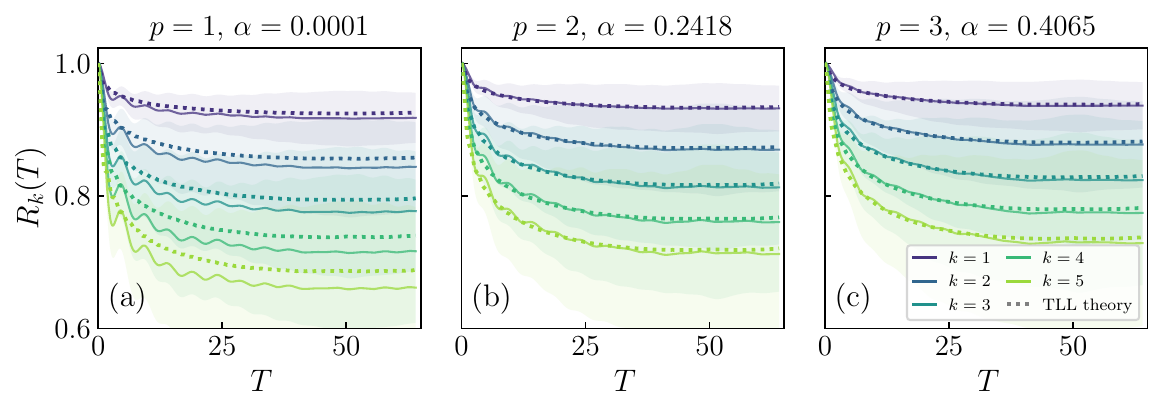}
\caption{
Comparison of filter orders $p=1,2,3$ for the XXZ chain with $L=64$, $\Delta=0.5$, and $\sigma_h=0.05$.
Solid curves are TEBD data and dotted curves are the TLL prediction from Eq.~\eqref{eq:main_result}, with $k=1,\dots,5$ from top to bottom in each panel.
The UV cutoff $\alpha$ is fitted from $k=1$ and held fixed for all $k$.
Increasing $p$ sharpens the suppression of the $2k_F$ disorder component and improves the theory and numerics agreement.
}
\label{fig:highorder_filter}
\end{figure}

As shown in Fig.~\ref{fig:highorder_filter}, the agreement between the TEBD numerics and the TLL prediction from Eq.~\eqref{eq:main_result} improves systematically with increasing $p$. In particular, the fitted UV cutoff $\alpha_{\rm fit}$ grows with $p$, reflecting the broader real-space support of the filter kernel and the reduced sensitivity to lattice-scale physics. The residual mean-squared deviation between theory and numerics decreases monotonically with $p$ across all tested values of $\Delta$ and $k$, confirming that higher-order filters bring the microscopic model closer to the continuum forward-scattering regime.

\section{Numerics}

\begin{figure}
    \centering
    \includegraphics[width=1.0\linewidth]{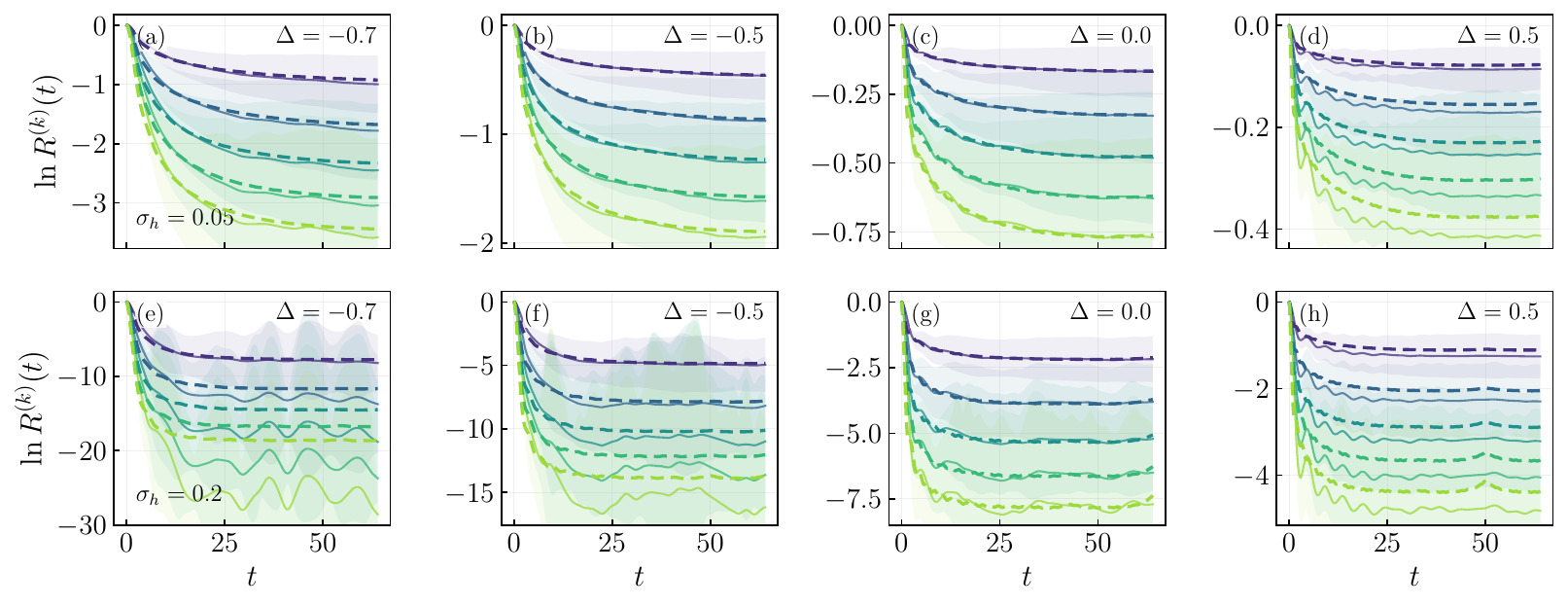}
    \caption{Time dependence of the normalized FP $R_k(T)$ for the interacting XXZ chain with $L=64$, $J=1$, $\Delta=-0.7,-0.5,0.0,0.5$, and $\sigma_h=0.05$ for panels (a)–(d) and $\sigma_h=0.2$ for panels (e)–(h). In each panel, different colors correspond to $k=1,\dots,5$ from top to bottom. Solid curves with shaded uncertainty bands are TEBD numerics. Dashed curves are the TLL prediction from Eq.~\eqref{eq:main_result}. For each $\Delta$, a single UV cutoff $\alpha$ is extracted from a $k=1$ fit in an intermediate time window and held fixed for all $k$.}
    \label{fig:S_FigmultifitFP}
\end{figure}

In this section, we present additional numerical results. In Fig.~\ref{fig:S_FigmultifitFP} we show the full numerical results corresponding to Fig.~(3) of the main text. We find that for $\sigma_h=0.2$, $\Delta=-0.7, -0.5$, the large-$k$ numerics deviate from the prediction, which we attribute to limited numerical precision arising from the small values involved. This also explains the deviation of the collapse curves at large $g$ in Fig.~(3) of the main text.

In addition, we present fits of $\alpha$ for $\sigma_h=0.05$, with all parameters summarized in Tab.~\ref{tab:ll_params}. We find that no positive value of $\alpha$ improves the agreement between the prediction and the numerics for $\Delta=0.7$, indicating the presence of backward-scattering effects.

\begin{table}[htbp]
  \centering
  \begin{tabular}{c c c c c c}
  \toprule
  $\sigma_h$ & $\Delta$ & $u$ & $K$ & $g$ & $\alpha_{\mathrm{fit}}$ \\
  \midrule
  0.05 & $-0.7$ & 0.4781 & 1.9749 & 5.50e-02 & $0.6500$ \\
  0.05 & $-0.5$ & 0.6495 & 1.5000 & 2.26e-02 & $0.4625$ \\
  0.05 & $+0.0$ & 1.0000 & 1.0000 & 6.37e-03 & $0.0799$ \\
  0.05 & $+0.5$ & 1.2990 & 0.7500 & 2.83e-03 & $\sim 0$ \\
  \bottomrule
  \end{tabular}
    \caption{Tomonaga-Luttinger liquid parameters and fitted UV cutoff $\alpha$ for the XXZ chain with $J=1$ at zero temperature. The dimensionless coupling is $g = 8\pi K \gamma / u^2$
   with $\gamma = \sigma_h^2/\pi^2$. The UV cutoff $\alpha$ is obtained by minimizing the mean-squared deviation between the $k=1$ TEBD data and Eq.~\eqref{eq:main_result} in the intermediate-time window $T\in[10,50]$, then held fixed for all $k$.}
  \label{tab:ll_params}
  \end{table}

\end{document}